# LBNL HIGH FIELD CORE PROGRAM*

S. Caspi Lawrence Berkeley National Laboratory, Berkeley, CA, USA

*Abstract*

The LBNL Superconducting Magnet Group mission is to develop and establish the technologies associated with high field superconducting magnets in order to provide cost-effective options for the next-generation high-energy physics accelerators. The research effort is part of the group core program and the development is part of the LARP program discussed elsewhere at this workshop.

## INTRODUCTION

In the past twenty years the LBNL core program has made the following contributions towards high field magnet using $Nb_3Sn$ conductor technology:

- Engineering properties of superconducting wires
- Cabling of traditional and advanced wires
- "Wind-and-React" coil fabrication technology
- Advance concepts for mechanical support
- New concepts for magnet assembly
- Modeling capabilities and diagnostic tools

The impact on the High Energy Physics community was the possible advance of a high energy/luminosity frontier of the LHC. The core program is focused on 1) conductor R&D and cable manufacturing, 2) magnet design, construction and testing and 3) new concepts and analysis.

## HIGH FIELD MAGNETS

Progress in the attainable dipole fields made with $Nb_3Sn$ conductor is plotted in Fig. 1. The type of magnets built and tested by LBNL varies from Cos-Theta (D20-13.8 T, 50 mm bore) to Common Coil (RD3-14.5 T) to Block (HD1-16 T) (Fig. 2). Other magnet were also built and tested as intermediate steps (Fig. 3)

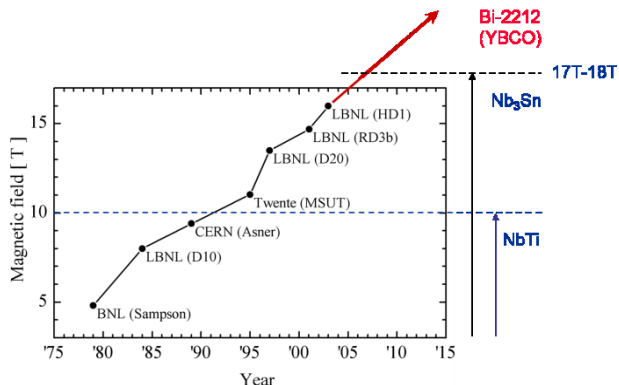

Figure.1: Progress of $Nb_3Sn$ dipole magnets.

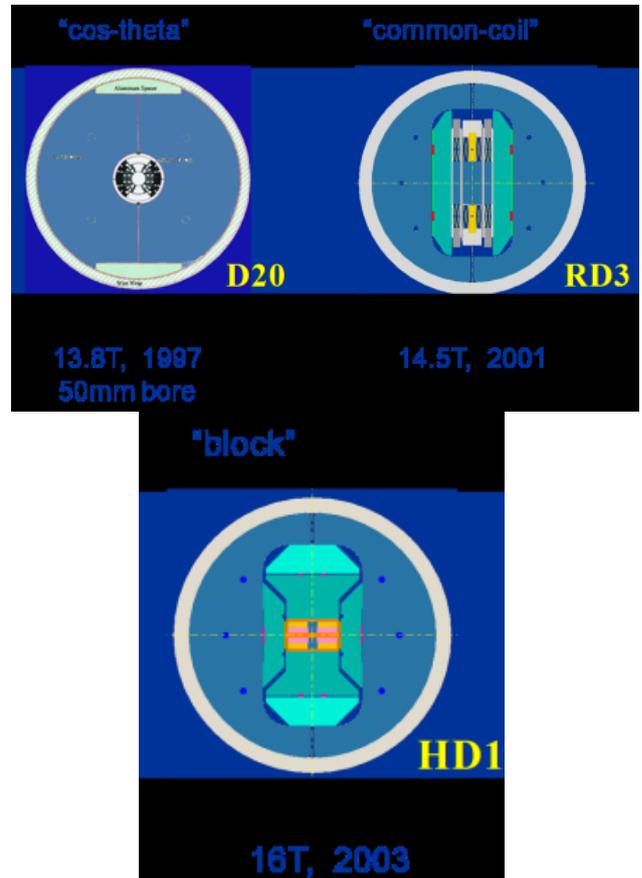

Figure 2: Three different configurations of dipole magnets constructed at LBNL

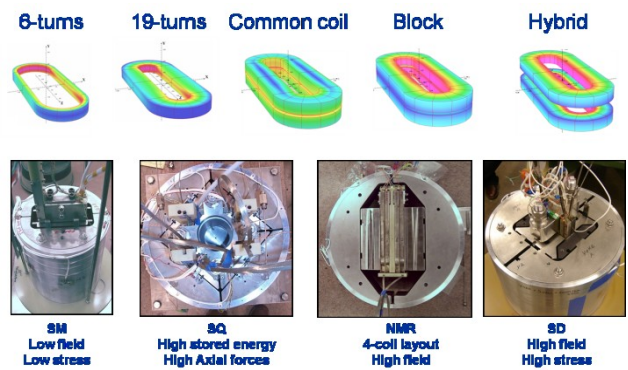

Figure 3: Left to right: SM- low field and low stress, SQ- high stored energy, high axial force, NMR-four coil layout high field, SD- high field high stress block design.

*Conductor Development*

The US HEP Conductor Development Program (CDP) has coordinated $Nb_3Sn$ work between National labs, universities and industry. Over the past twenty years the program main achievements were the doubling of $Nb_3Sn$ current density and the improvement of wire uniformity and piece length. The program continues to work on improving the current density of $Nb_3Sn$ as well as reducing its sub-filament size (Fig. 4). From 2007 the CDP supported the development Bi-2212 demonstrating its performance in simple configurations.

To understand the relations of the conductor state at different scales, a hierarchical model of the strain state has been developed. The model included nonlinear properties and enabled computing the strain at the filament level including stress in micro-scales due to macro loading and nonlinear deformation. The work also included cool-down effects. The work has been extended to the manufacturing and optimization of cables (Fig. 5).

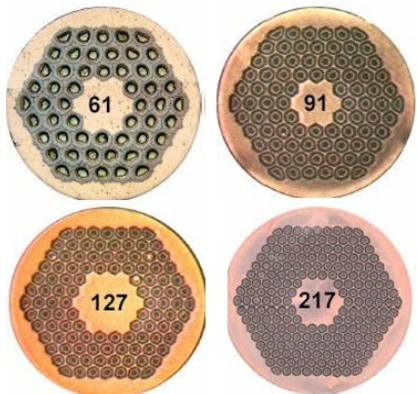

Figure 4: $Nb_3Sn$ strands with different filament number

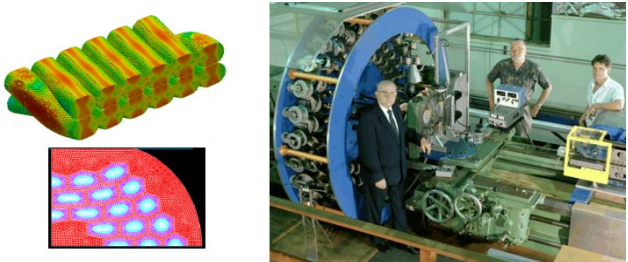

Figure 5: Manufacturing and analysis of cables

## Magnet Development

The manufacturing of magnets with $Nb_3Sn$ coils requires the integration between CAD, analysis and manufacturing (Fig. 6). The process of winding and curing coils using metallic parts (Fig. 7) reaction at 650°C, instrumentation and impregnation has been made into a continuous integrated process that closely interacts with analysis. Magnetic and structural analysis follows the magnet design from its room temperature assembly and pre-stress through cool-down and excitation to "short-sample". The magnet assembly uses "key and bladder" technology and the final pre-stress is reached during cool-down mainly due to the thermal expansion difference between iron and aluminium (Fig. 8-9).

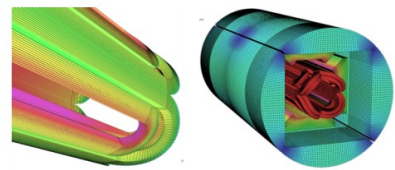
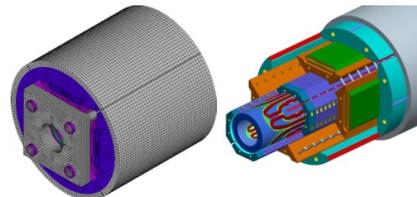

Figure 6: Integrated design between CAD, magnetic and structural analysis

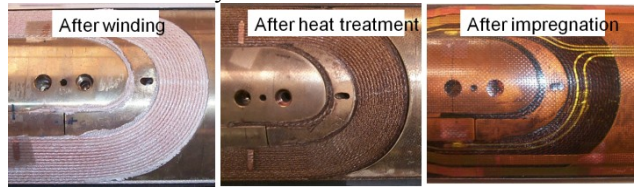

Figure 7: View of magnet "end" at different stages.

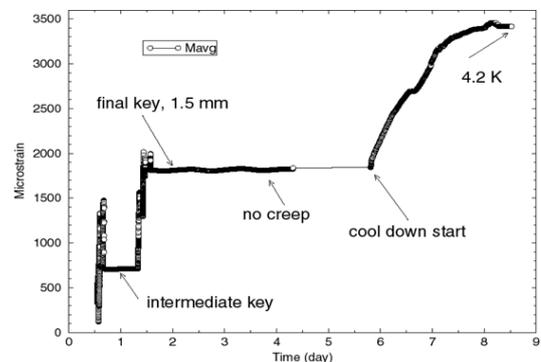

Figure 8: Typical increase of shell tension during assembly and cool-down

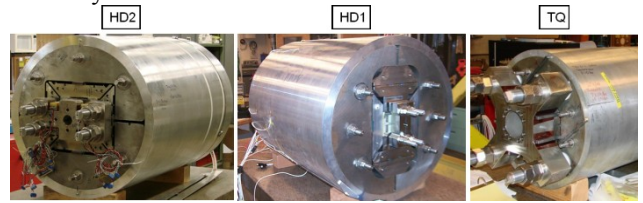

Figure 9: Aluminium shells used to pre-stress coils

As part of the development of $Nb_3Sn$ high field magnets for the next generation of HEP colliders [1] the LBNL Superconducting Magnet Program is fabricating and testing a series of $Nb_3Sn$ dipoles magnet HD2/3 (see Fig. 10-11). References on the conceptual design, the coil and structure mechanical analysis, the fabrication and assembly procedure and the field quality expectations are in [2-4]. Results of several tests, carried out at the LBNL test facility, are shown in Figure 12 indicating low and incomplete training curves. Most quench origins were located at the end of the straight section just prior to the start up of the bend. A subsequent autopsy at that location showed an unintended step in the upper block (Fig. 13) created by the cable hard-way bend. In HD3 coils, under construction, the radius of the bend was increased to ease the bend and a supporting "membrane" was added between layers. Other test results including strain gauges measurements, training performance, quench locations, and ramp-rate studies are reported in [5]. Other improvements now include curing of coils (using a binder) to better position them prior to reaction. By reducing the reaction temperature of HD3 coils, a more conservative approach was taken by a corresponding reduction of the current density from 3300 A/mm$^2$ (12 T, 4.2 K) in HD2 to 3000 A/mm$^2$ in HD3. The impact of all such changes reduced the short-sample bore field from 15.6 T in HD2 to 14.9 T in HD3.

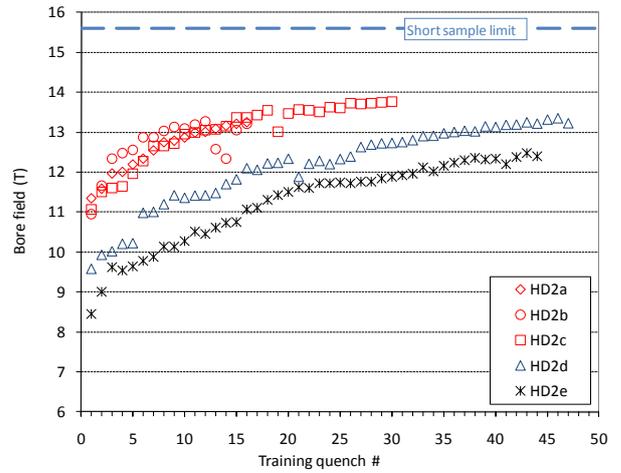

Figure 12: Bore field (T) as a function of training quenches. The short sample limit of 15.6 T bore field corresponds to a coil peak field of 16.5 T.

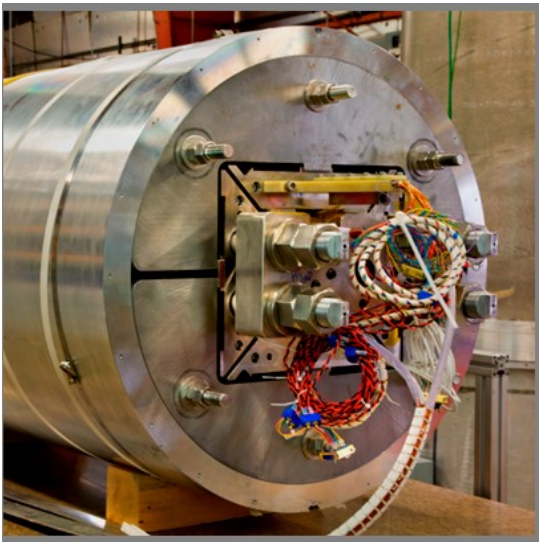

Figure 10: Magnet HD2

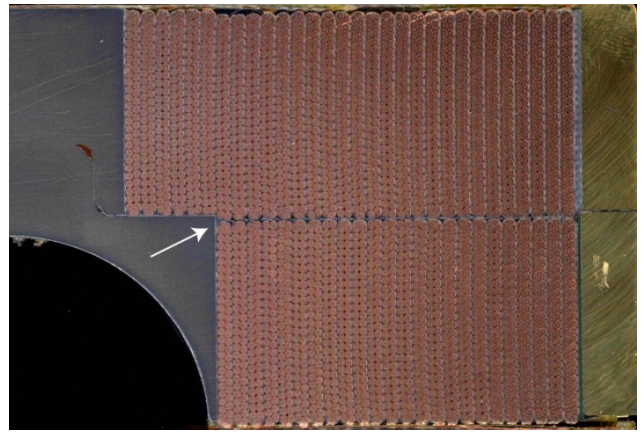

Figure 13: Cross-section cuts of HD2 coil #1 close to the beginning of the hard-way.

*Analysis*

LBNL has been developing 3D finite element models to predict the behaviour of high field Nb$_3$Sn superconducting magnets [6]. The models track the coil response during assembly, cool-down and excitation with particular interest on displacements when frictional forces arise. As Lorentz forces can be cycled and irreversible displacements can be computed and compared with strain gauge measurements. Analysis on the release of local frictional energy during magnet excitation results in a temperature increase that can be calculated. Magnet quenching and training is then correlated to that level [7]. Figures 14-15 show the results of the analysis using the programs TOSCA and ANSYS.

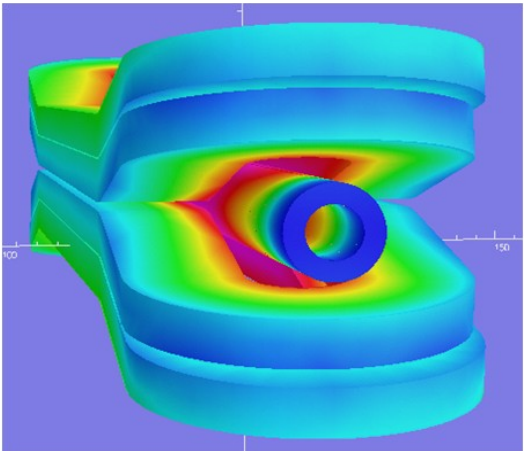

Figure 11: Computed field magnitude of HD2

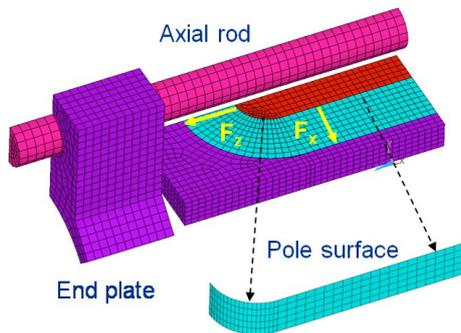

Figure 14: View of coil, island, end plate, axial support rod and contact surface between coil and island.

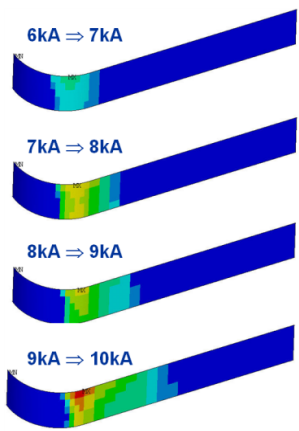

Figure 15: Surface between coil and island showing a potential increase in energy release at higher currents a leading cause of training.

## ACKNOWLEDGEMENT

The Superconducting Group Team:

D. Arbelaez, B. Bingham, S. Caspi, D. Cheng, B. Collins, D. Dietderich, H. Felice, A. Godeke, R. Hafalia, J. Joseph, J. Krishnan, J. Lizarazo, M. Marchevsky, S. Prestemon, G. Sabbi, C. Vu, X. Wang. P. Bish, H. Higley, D. Horler, S. King, C. Kozy, N. Liggins, J. Swanson, P. Wong D. Pickett, J. Smithwick, G. Thomas, K. Miho

## REFERENCES

[1] A. F. Lietzke, et al., "Test results for HD1, a 16 Tesla $Nb_3Sn$ dipole magnet", IEEE Trans. Appl. Supercond., vol. 14, no. 2, pp. 345-348, (June 2004).

[2] G. Sabbi, et al., "Design of HD2: a 15 T $Nb_3Sn$ dipole with a 35 mm bore", IEEE Trans. Appl. Supercond., vol. 15, no. 2, pp. 1128-1131, (June 2005).

[3] P. Ferracin, et al.,"Mechanical design of HD2, a 15 T $Nb_3Sn$ dipole magnet with a 35 mm bore", IEEE Trans. Appl. Supercond., vol. 16, no. 2, pp. 378-381, (June 2006).

[4] P. Ferracin, et al., "Development of the 15 T $Nb_3Sn$ Dipole HD2", IEEE Trans. Appl. Supercond., vol. 18, no. 2, pp. 277-280, (June 2008).

[5] P. Ferracin, et al., "Assembly and Test of HD2, a 36 mm Bore High Field $Nb_3Sn$ Dipole Magnet", IEEE Trans. Appl. Supercond., vol. 19, no. 3, pp. 1240-1243, (June 2009).

[6] S. Caspi, et al., "Towards integrated design and modeling of high field accelerator magnets", IEEE Trans. Appl. Supercond., vol. 16, no. 2, pp. 1298-1303, (June 2006).

[7] P. Ferracin, S. Caspi, and A.F. Lietzke, "Towards Computing Ratcheting and Training in Superconducting Magnets", IEEE Trans. Appl. Supercond., vol. 17, no. 2, pp. 2373-2376, (June 2007)